# Realismo e metafísica na mecânica quântica

Raoni Wohnrath Arroyo e Jonas R. Becker Arenhart[1]

## Resumo

De acordo com o realismo científico, a ciência nos dá uma descrição aproximadamente verdadeira de como é o mundo. Mas o que isso significa? Neste capítulo, focamos nos aspectos ontológicos e metafísicos dessa discussão. Isto é, nos preocupamos com as seguintes questões: o que existe, de acordo com nossas melhores teorias científicas? E como são essas coisas que as teorias dizem existir? Partimos do pressuposto de que um realismo científico genuíno deve lidar com essas questões. Tomando a mecânica quântica não relativista como estudo de caso, discutimos sobre alguns desafios atualmente enfrentados por uma postura realista genuína. Argumentamos, primeiramente, que no aspecto ontológico, realistas encontram-se no mesmo barco que a ciência, isto é, sem justificativa epistêmica suficiente para adotar a crença nas entidades postuladas por uma só teoria quântica, haja vista que a experiência (atualmente) não é capaz de decidir entre teorias rivais; no aspecto metafísico, realistas também encontram problemas com a escolha de teorias, dada a possibilidade de associar mais de uma teoria metafísica com a descrição das entidades postuladas por cada teoria quântica. O mesmo vale para as alternativas estruturalistas, já que, mesmo que aceitemos que as teorias científicas se comprometem com a existência de estruturas, mas não de objetos, ainda não sabemos o que são, metafisicamente, estruturas. Por fim, avançamos na discussão com o método meta-Popperiano, que serve para reduzirmos as alternativas metafísicas associadas aos postulados ontológicos das teorias científicas. Podemos não saber qual é a alternativa correta, mas conseguimos checar progressivamente quais alternativas metafísicas são incompatíveis com cada teoria científica.

## 1.Introdução

Tradicionalmente, admite-se que ser "realista" sobre algo ou sobre algum domínio do conhecimento consiste em manter algum tipo de crença na *existência independente* de algumas entidades nesse domínio ou campo. Por exemplo, se você disser ser realista em relação ao mundo externo, é muito provável que você queira dizer que *acredita na realidade* do mundo externo, isto é, que você não é um cérebro em uma cuba, por exemplo. Essa última situação envolveria adoção da tese contrária, que nega a realidade do mundo externo. Este capítulo trata do realismo científico, de modo que a pergunta que mais nos interessa aqui é a seguinte: *o realismo científico requer crença sobre o quê*?

Comecemos pelo começo. Uma das principais tarefas da filosofia da ciência é buscar compreender a natureza do empreendimento científico. Uma das respostas é o realismo científico, grosso modo caracterizado pela afirmação de que a ciência descreve corretamente (pelo menos aproximadamente) como o mundo é, não apenas em relação aos aspectos observáveis, mas também aos inobserváveis — e os diferentes tipos de antirrealismo negam isso. Por uma questão de simplicidade, daqui pra frente desconsideraremos a parte "científica" dos termos. Isto é, mesmo que o assunto aqui seja o realismo científico, os termos "realismo científico" e "realismo" serão empregados de forma intercambiável.

Segundo o realismo, as nossas melhores teorias científicas fornecem uma imagem razoavelmente fiel de como o mundo realmente é. Nessa perspectiva, as teorias científicas revelam a realidade: elas descrevem-na verdadeiramente — ou ao menos *aproximam-se* de uma descrição verdadeira. Ao contrário, as diferentes formas de antirrealismo negam que a ciência nos forneça uma descrição aproximadamente verdadeira da realidade. Isso pode ser desdobrado de diferentes maneiras, seja como sugerindo que as nossas melhores teorias não são mais do que simplesmente *empiricamente adequadas*, sem se comprometer com sua verdade, seja através de afirmações mais radicais de que nossas melhores teorias são apenas instrumentos para previsões e controle do nosso contorno.

Naturalmente, ao sustentarem que nossas melhores teorias são aproximadamente verdadeiras, realistas se comprometem, em particular, com as afirmações existenciais que essas teorias acabam fazendo. De fato, teorias científicas parecem fazer uso de entidades específicas para que suas explicações funcionem. Quando as teorias científicas se referem a coisas que podem ser mostradas diretamente a qualquer pessoa dotada de faculdades perceptivas plenas, como objetos fazendo parte da mobília de um quarto, de modo geral, realistas e antirrealistas têm pouco ou nada a discordar. O debate mais acalorado se dá quando as teorias tratam sobre coisas que não são acessíveis nesse sentido. Desse modo, trata-se de um debate que depende da definição do termo "observável": somente coisas que são percebidas diretamente pelos sentidos humanos contam como "observáveis"? Ou seria legítimo caracterizar como "observáveis" aquelas coisas sobre as quais temos acesso por meio de instrumentos? Em caso afirmativo, qual seria um grau aceitável da complexidade desses instrumentos?

Como se sabe, o termo "observável" é vago. Ou seja, seu campo de aplicação não é claro. Isso poderia, a princípio, dificultar o debate, mas não precisamos nos deter nesse aspecto da discussão aqui. Existem situações limite, nas quais a noção se aplica suficientemente bem: é seguro dizer, por exemplo, que um planeta visível sem auxílio de instrumentos é observável. Da mesma forma, é seguro dizer que um objeto subatômico, como um elétron, não é. Assim, mesmo que não se tenha fronteiras claras para o que conta como observável ou não, a discussão pode prosseguir, pois temos garantido que o termo separa duas classes de entidades.

Com essa polêmica no lugar, um ponto central que distingue as abordagens realista e antirrealista refere-se ao estatuto ontológico das entidades envolvidas nas nossas teorias de maior sucesso: seria legítimo acreditar na existência de entidades não observáveis postuladas nessas teorias? Enquanto o realismo sustenta que essas entidades existem, apesar de nosso conhecimento ou capacidade de detectá-las, o antirrealismo, por outro lado, mantém uma atitude cética em relação ao assunto.

Então aqui podemos trazer para o centro do palco a principal questão de disputa entre realistas e antirrealistas que vai nos ocupar neste capítulo: um dos principais aspectos de desacordo entre realistas e antirrealistas concerne à disputa acerca da nossa crença na *existência* de entidades não-observáveis postuladas em nossas melhores teorias científicas. A disputa concerne a nossa capacidade de justificar nosso assentimento à crença na existência dessas entidades. Esse é um problema epistemológico, relacionando nossas melhores descrições do mundo com a própria realidade. Como veremos, realistas buscam criar o vínculo entre teorias e realidade, enquanto que antirrealistas buscam sustentar que o sucesso empírico não requer a crença na existência das correspondentes entidades.

Dito isso, outra questão imediatamente emerge: no que realistas acreditam? Isto é, qual o conteúdo de sua crença? Trata-se de uma questão que, para algumas pessoas, se afigura fundamental. Crer na existência de determinadas entidades não-observáveis parece requerer que sejamos capazes de dar uma descrição bastante clara de como são essas entidades. É aqui que diferentes formas de realismo podem surgir. Algumas vão sugerir que a ciência encarna nossa melhor forma de acesso ao mundo, e a sua descrição, consequentemente, é suficiente para formar uma imagem de mundo. Outras vão sugerir que uma camada metafísica deve ser adicionada sobre as entidades postuladas pela ciência, para que a imagem de mundo seja completa e clara o suficiente. Como veremos, nesse contexto ampliado, um dos principais desafios para realistas será manter sua fé na capacidade epistêmica da ciência e ainda garantir uma imagem clara de mundo.

Iniciaremos nossa investigação na seção 2 apresentando os argumentos centrais no debate 'realismo *vs* antirrealismo' em filosofia da ciência. Na seção 3, o debate é levado para o domínio quântico, e ilustramos como o compromisso realista se vincula com as chamadas interpretações da teoria. Conforme argumentaremos, o compromisso realista está diretamente relacionado com a ontologia de uma interpretação da teoria, de modo que a tese realista é vista como uma tese *ontológica* sobre a ciência. Na seção 4, a caracterização do realismo sobre a ciência é descrita em ainda mais detalhes uma vez que, segundo um tipo de realista, a imagem de mundo obtida através da ciência deve ser complementada com uma descrição *metafísica*: somente assim teríamos uma imagem clara o suficiente da realidade para ser digna do título de realismo. Analisaremos como essa atribuição de uma roupagem metafísica pode ser feita, e como realistas enfrentam dificuldades nesse aspecto particular. Um exemplo da discussão acerca da metafísica da individualidade na mecânica quântica (MQ) ilustra o problema. Em particular, dificuldades vistas na seção 2 reaparecem com uma nova roupagem e enfraquecem a posição realista. Por fim, nos beneficiando do levantamento dos principais obstáculos enfrentados pelo realismo, delineamos uma saída para realistas que desejam vestir metafisicamente sua ontologia, e ainda assim, ter algum benefício da ciência nesse processo. Isso gera uma forma de metametafísica associada com a ciência, e ilustra como a metafísica pode se integrar positivamente com a ciência. Concluímos na seção 5.

## 2.Realismo e antirrealismo: Degustando os principais argumentos

Nesta seção, destacamos três argumentos principais na literatura a respeito do debate entre essas duas concepções filosóficas (um guia excelente e mais detalhado por esses argumentos pode ser encontrado em French, 2009, cap. 7 e 8). Iniciaremos com um argumento bastante aprazível ao realismo, chamado de "argumento dos milagres" (às vezes referido como "argumento sem milagres", já que milagres não são bem quistos em explicações). É considerado um dos argumentos mais fortes em favor do realismo científico. Em seguida, apresentaremos um argumento com notas agridoces, com a metaindução pessimista, e fecharemos com o amargor da subdeterminação.

### 2.1.O argumento dos milagres

Um dos principais argumentos em defesa do realismo científico reside na explicação da adequação empírica das teorias científicas. Uma atitude intuitiva em relação ao sucesso das teorias científicas ao fazerem descrições e predições é considerar que essas coisas dão certo *porque* essas teorias falam diretamente sobre o mundo, portanto são teorias *verdadeiras* num sentido correspondentista da verdade. Não fosse esse o caso, o fato de uma teoria científica fazer previsões corretas seria uma mera coincidência, e a adequação empírica dessas previsões seria considerada um milagre.

Além disso, seria um milagre constante para toda previsão concebível com sucesso experimental. Como Putnam (1975, p. 73) famosamente declarou, o realismo científico é "[...] a única filosofia que não faz do sucesso da ciência um milagre".[2] Assim, o realismo tem um argumento positivo: o argumento dos milagres. Esse é um argumento poderoso, uma vez que a ciência, como um fenômeno cultural, tem moldado visões de mundo e condições materiais da vida humana.

No que diz respeito à ontologia, isto é, em relação às questões de *existência*, o argumento dos milagres aceita o ponto de sucesso empírico como uma condição suficiente para que as entidades existentes na teoria sejam consideradas entidades existentes no mundo. Ou seja, o argumento diz que os objetos postulados pelas teorias científicas devem corresponder aos objetos do mundo, a menos que aceitemos sem reservas a ideia de que as aplicações da ciência sejam um milagre.

2No original: "[...] the only philosophy of science that doesn't make the success of science a miracle".

De fato, esse é o único argumento tradicionalmente aceito como sendo a favor do realismo científico. Os outros dois argumentos que examinaremos adicionam notas mais amargas à degustação realista.

**2.2 Metaindução pessimista**

De uma perspectiva histórica, várias teorias científicas obtiveram sucesso empírico em suas previsões e, no entanto, foram substituídas por outras teorias que apresentaram maior sucesso empírico e/ou maior domínio de explicação. Ou seja, a história da ciência reconhece como falsas inúmeras teorias científicas que antes eram consideradas verdadeiras.

O argumento é: o fato de nossas teorias científicas serem bem-sucedidas atualmente não garante seu sucesso no futuro; de fato, a história da ciência nos fornece boas razões para não acreditarmos que isso ocorrerá. Quer dizer, ao olharmos para trás na história da ciência, poderemos ver, dentre outras coisas, uma lista de *ontologias abandonadas*, isto é, de entidades que já foram tidas como *reais*, e que contribuíram *de fato* para a economia de teorias científicas em determinado período histórico, mas que acabaram sendo descartadas, dando lugar para novas ontologias associadas a teorias científicas sucessoras. Exemplos disso variam desde elementos químicos sutis, como o éter, até coisas maiores e mais pesadas, como o planeta Vulcano (uma lista recheada de exemplos desse tipo pode ser encontrada em Laudan, 1981, p. 33, e em French, 2009, p. 97). Esse é o chamado argumento da "metaindução pessimista". É um raciocínio indutivo (*à la* Hume), que emprega a história da ciência ao falar sobre a ciência (sendo, portanto, *meta*indutivo), ao passo que mantém uma atitude cética quanto à possibilidade de estarmos à beira de uma explicação final na ciência (portanto *pessimista*).

Como o foco de nossa discussão reside na existência das entidades com as quais as teorias científicas estão comprometidas, o argumento apresentado pela metaindução pessimista é o seguinte: a substituição e o subsequente abandono das teorias científicas resultam na substituição das entidades com as quais a teoria estava existencialmente comprometida. Portanto, as entidades postuladas pela teoria não devem ser estendidas para uma existência no mundo real, ou seja, os objetos associados às teorias científicas não devem ser tomados como existindo fora das teorias que as introduziram.

Embora o aspecto histórico do argumento seja altamente intuitivo, ele não tem boas garantias em relação à crítica ao realismo científico. Isso porque é um argumento que apresenta um apelo subjetivista: o pessimismo. Repare: alguém poderia manter uma atitude *otimista* em relação à perspectiva histórica, concedendo erros do passado e defendendo opiniões como "*Agora sim* estamos prestes a obter a verdade" ou "*Agora sim* nós descobrimos as entidades fundamentais do universo". Em última análise, a postura otimista carece de justificativas tanto quanto a postura pessimista — afinal, é o problema da indução que está em jogo aqui. Notas agridoces. Quer dizer, não é como se fosse apenas um probleminha: quatro séculos depois, e ainda não conseguimos achar razões *necessárias* para a justificação de raciocínios indutivos; conseguimos, no entanto, manter uma discussão saudável em ciência *mesmo se* considerarmos o problema de Hume um problema filosófico, como tantos outros, insolúvel. Passemos à sobremesa.

**2.3.Subdeterminação**

Quando temos (ao menos duas) alternativas igualmente bem-sucedidas no domínio empírico, não podemos apelar para nenhum dado do domínio que nos permitiria escolher entre teorias rivais. Assim, é dito que os dados empíricos não determinam a teoria que os explica, o que é a mesma coisa que dizer que "as teorias são subdeterminadas pelos dados", ou que há uma "subdeterminação da teoria pelos dados". Muller (2011, p. 230) organiza ainda mais o assunto da subdeterminação, chamando essa subdeterminação teórica de "primeira tese da subdeterminação", em distinção à subdeterminação metafísica, chamada de "segunda tese da subdeterminação", que ficou muito famosa com o trabalho seminal de French e Krause (2006) e os estudos subsequentes em relação à metafísica da MQ.

A MQ tem sido um exemplo notável disso. Existem várias maneiras diferentes de interpretar os mesmos fenômenos no nível quântico. Considerando apenas as evidências empíricas, no entanto, os resultados das previsões de fenômenos e suas aplicações tecnológicas são mantidos. Isso sugere que a MQ, como teoria científica, não fornece os elementos necessários para decidir qual interpretação está correta. Portanto, se a interpretação da MQ é subdeterminada por dados experimentais, como alguém pode afirmar que uma interpretação em particular é uma descrição real do mundo, enquanto outra é uma mera construção? Em outras palavras, não parecemos ter justificativas suficientes para sermos realistas sobre uma interpretação em vez de outra.

Do ponto de vista ontológico, o argumento contra o realismo científico reside na dificuldade de afirmar que as entidades de uma interpretação existem enquanto as entidades de outras não, uma vez que, do ponto de vista empírico, as alternativas são equivalentes. Isto é, cada uma dessas soluções povoa a teoria com um mobiliário diferente. E, ao que parece, não temos critérios suficientes para afirmar qual delas corresponde ao mobiliário do mundo.

O amargor prometido leva algum tempo para se desenvolver no paladar. Ele ainda não é evidente. Afinal, isso tudo parece ser uma questão que não sai do âmbito da epistemologia, não chegando à ontologia: "nós não *sabemos* qual a certa", você poderia dizer, "muito embora *haja* uma". O retrogosto começará a aparecer na próxima seção.

**3..Realismo no domínio quântico**

Tradicionalmente, admite-se que *a MQ deve ser interpretada por causa do problema da medição*. Somente na sentença anterior, podemos identificar ao menos três questões terminológicas e metodológicas, as quais comentaremos brevemente antes de prosseguir.

Em primeiro lugar, destacamos que há um caloroso debate acerca da adequação do termo "interpretação". Há quem considere que cada "interpretação" da MQ é uma interpretação em um sentido intuitivo, que *atribui sentido* à MQ (cf. Jammer, 1974), mas há dissidência: algumas pessoas afirmam que cada interpretação é, por si, uma *teoria física* madura (cf. Maudlin, 1995; Ćirković, 2005); outras, que somente *alguns* casos são teorias quânticas diferentes, como é o caso da mecânica Bohmiana (cf. Bohm, 1951). A discussão que propomos aqui, no entanto, independe da adoção de uma posição nesse debate terminológico acerca de como entender propriamente uma interpretação. Sem perda de generalidade ou precisão conceitual, podemos dizer que a *primeira tese da subdeterminação* é exemplificada pelo problema da medição na MQ, e que representa um obstáculo para a adoção de um realismo científico, *mesmo que* cada solução do problema da medição seja considerada uma *teoria física* diferente ou *interpretações* diferentes de uma (única) teoria física: no final do dia pode haver somente uma alternativa para representar corretamente os fenômenos físicos em questão.

Em segundo lugar, há uma confusão terminológica sobre o termo "medição". Um dos tópicos mais importantes na filosofia da ciência é definir o que é uma medição, e os esforços para tal são discutidos nas "teorias da medição". No entanto, esse debate é circunscrito pelas noções de "grandezas", seus aspectos definidores, quantitativos, qualitativos, e a metodologia para que possamos atribuir, de maneira precisa, valores a coisas — por exemplo, uma (dentre várias) questões da "filosofia da medição" debruça-se sobre os desafios que surgem quando queremos "medir", num sentido de relacionar estruturas matemáticas, numéricas, a objetos físicos (cf. Pinheiro, 2018, 2019). O ponto é: esse debate não se relaciona com o conteúdo das chamadas "teorias da medição" no contexto específico do problema da medição na MQ — que, por sua vez, é um dos tópicos mais debatidos acerca dos fundamentos da MQ. Cientes dessa confusão, podemos avançar sem contaminarmos essa discussão (suficientemente controversa) com maiores mal-entendidos: neste capítulo, estamos dialogando *exclusivamente* com a literatura que trata do chamado "problema da medição" na MQ — ainda que o termo "medição" possa estar sendo empregado de modo impreciso (e provavelmente está).

Por fim, as motivações para a interpretação variam. Por exemplo, há quem considere que, sem uma resposta ao problema da medição, a MQ é *inconsistente* (cf. Maudlin, 1995), que torna-se *trivial* (cf. Lewis, 2016, p. 50), ou que esteja envolvida em uma *contradição empírica* (cf. Ruetsche, 2018, p. 296); há, ainda, quem considere que a MQ, antes da interpretação, não é nem mesmo uma teoria física, mas uma *proto*-teoria (cf. de Ronde, 2018), ou um *esquema de postulados metalinguísticos*, que podem ser transformados em axiomas específicos para a construção de teorias quânticas (cf. Arroyo, 2020, cap. 2), ou, ainda, um *framework* teórico sobre o qual teorias físicas concretas possam ser expressas (cf. Wallace, 2020).

Como veremos, não precisamos ir tão a fundo a ponto de axiomatizar a MQ, nem buscar uma real inconsistência em seu formalismo para motivar a análise do problema. O ponto em comum a todas essas motivações é: temos um problema com algumas assunções básicas que são feitas sobre a MQ. É, no sentido etimológico do termo, um *paradoxo*. O termo "paradoxo", da etimologia grega, denota algo que está além da opinião aceita em geral. Trata-se basicamente de um problema com premissas consideradas razoáveis, mas em que ao menos uma delas deve ser rejeitada. Para mostrarmos quais são essas assunções cuja conjunção é problemática, utilizaremos a taxonomia oferecida por Maudlin (1995), que é a maneira tradicional de enunciar o problema.

Um último comentário antes de passarmos à exposição e discussão do problema propriamente: nosso intuito é que a presente discussão seja amigável às pessoas interessadas nos aspectos filosóficos dessa discussão que envolve a MQ como exemplo, mas que têm pouco ou nenhum treino nas artes Jedis da física. Por isso, muitas vezes trocaremos a precisão pela clareza quando tocarmos nos assuntos filosóficos que estão em jogo na física. Exames mais rigorosos (mas ainda inclinados à filosofia) sobre o problema da medição e o formalismo da MQ podem ser encontrados, respectivamente, em Pessoa Junior (1992) e Krause (2016; 2017). Além disso, em um tom mais técnico, trabalharemos exclusivamente com formulações da MQ envolvendo espaços de Hilbert e funções de onda, já que são abordagens dominantes na filosofia da física. Para formulações da MQ que não envolvem espaços Hilbert, ver Styer *et al*. (2002); para uma formulação da MQ sem funções de onda, ver Schiff e Poirier (2012).

Teorias quânticas geralmente usam uma representação matemática chamada "função de onda" para descrever um estado físico de um sistema. Uma maneira bastante elementar de enunciar o problema de medição é através de um simples questionário sobre essa função de onda:

1. A função de onda é uma descrição *completa* do estado físico do sistema?
2. A função de onda sempre evolui de maneira *linear* e determinística ou, às

vezes, é descrita de alguma outra maneira?

3. A função de onda tem resultados *determinados*?

Considere a seguinte situação, tipicamente quântica: a função de onda $\psi$ descreve um sistema físico em uma situação que admite duas soluções distintas, digamos, $\psi_1$ e $\psi_2$. Se respondermos afirmativamente à primeira questão da lista, temos que todos os aspectos físicos do sistema em questão são descritos pela função de onda $\psi$. Por consequência da *linearidade*, a descrição da função de onda é: $\psi = \psi_1 + \psi_2$. Se também dissermos "sim" à questão do segundo item, então essa é a única evolução possível desse sistema físico. Agora, se também respondemos afirmativamente à terceira questão, temos um problema: a equação que descreve completamente o sistema físico não tem um único resultado determinado (ou seja, não resulta *exclusivamente* em $\psi_1$ ou $\psi_2$), e não temos outra evolução possível. *Voilà*! Eis o problema da medição. O que se chama de "interpretação" da MQ é uma solução para esse problema e, de acordo com essa formulação, isso requer que *ao menos uma* dessas questões deva ser respondida negativamente.

A primeira questão foi o palco de um dos debates mais acalorados na física desde Leibniz-Clarke, protagonizado principalmente por Niels Bohr e Albert Einstein nos anos 1930, mas que envolveu diversas outras pessoas (para mais detalhes sobre esse debate, ver Jammer, 1974 e Chibeni, 1997). Alguém que responda "não" a ela, diz, implicitamente, que dois sistemas descritos pela mesma função de onda são dois sistemas diferentes — isto é, *algo* está escapando à descrição da função de onda: ela não descreve, portanto, *completamente* os estados físicos do sistema em questão. A *teoria quântica padrão* ou a *interpretação de Copenhague* sustenta que a função de onda é completa. As soluções que sustentam que a função de onda não é completa postulam *variáveis ocultas* para suplementar essa função de onda. Vamos chamar essa solução de $MQ_V$. Exemplos de interpretações desse tipo são as interpretações dos conjuntos estatísticos (cf. Ballentine, 1970) e a teoria da onda piloto, também chamada de "mecânica Bohmiana" (cf. Bohm, 1951; 2014).

A questão 2 diz respeito à evolução da função de onda no tempo. Responder negativamente à questão 2 implica em acrescentar uma lei dinâmica não-linear no comportamento da função de onda. Essa lei dinâmica adicional é chamada de "colapso" ou "redução". Chamaremos soluções desse tipo de $MQ_C$. De acordo com uma teoria quântica com colapso, a função de onda $\psi$ é imediatamente transformada em um dos resultados possíveis em determinadas situações (isto é, *exclusivamente* $\psi_1$ ou $\psi_2$). A interpretação de Copenhague (cf. Jammer, 1974) e a teoria da "redução espontânea" (cf. Ghirardi, Rimini e Weber, 1986) estão dentre os exemplos desse tipo de solução.

Finalmente, a questão 3 diz respeito aos resultados da descrição física através da função de onda. Para quem quer que afirme que a função de onda é completa e nega que haja colapsos, só resta uma alternativa: negar que haja resultados únicos para uma medição. Em teorias desse tipo, a solução da nossa equação simples acima não é $\psi_1$ nem $\psi_2$, mas algo além. Inicia-se um experimento com apenas uma função de onda $\psi$, e termina-se com duas: uma resultando em $\psi_1$ e outra resultando em $\psi_2$. Soluções desse tipo são chamadas de "estado relativo", e chamaremos aqui de $MQ_R$, dentre as quais podemos destacar a mecânica quântica Everettiana (cf. Everett, 1957), e a teoria dos muitos mundos (cf. DeWitt, 1971; ver também o Capítulo 6 deste volume).

Considerando os argumentos a favor e contra o realismo científico esboçados na seção anterior, agora é possível abordar o quadro geral: como alguém pode adotar uma posição realista no que diz respeito à MQ? Considere o argumento positivo do realismo científico, o *argumento dos milagres*. Ruetsche (2015, seção 3.3) apresenta-o da seguinte forma:

1. A teoria *T* é bem-sucedida

2. A verdade de *T* é a melhor explicação para esse sucesso.
∴ *T* é verdadeira.

Essa esquematização do argumento mostra suas raízes abdutivas, que já foram criticadas anteriormente (cf. Fine, 1986, p. 115). Existem também versões do argumento dos milagres que não dependem de valores de verdade (cf. van Fraassen, 1980, p. 40). Nesse ponto do debate, é interessante deixar de lado essas linhas de crítica e prestar atenção a um significado possível do argumento dos milagres, apresentado por Ruetsche (2015, seção 3.3), considerando que ele se aplica a uma teoria científica específica, como a MQ:

1. A MQ é bem-sucedida
2. A verdade da MQ é a melhor explicação para esse sucesso.
∴ A MQ é verdadeira.

Como a MQ é uma das teorias empíricas mais bem-sucedidas até agora, a primeira premissa é garantida (Ruetsche, 2018; da Costa, 2019). Além disso, o afastamento deliberado de críticas a respeito de raciocínios abdutivos (ou da "inferência pela melhor explicação") e de argumentos guiados por valores de verdade permite aceitarmos a seguinte conclusão: a MQ é *verdadeira*. Nesse sentido, o seguinte argumento defendido por Ruetsche (2015, seção 3.3) é de grande interesse para nós neste capítulo: em *que* alguém acredita quando acredita na verdade de uma teoria? A resposta é:

> O que uma realista acredita, quando ela acredita em uma teoria *T* é uma *interpretação de T*, uma descrição de como possivelmente seria o mundo de acordo com *T*. [...] Uma interpretação da MQ diz à realista sobre a MQ *o que* ela acredita quando ela acredita na MQ. (Ruetsche, 2018, p. 293, ênfase original)[3]

De acordo com esse raciocínio, acreditar na verdade da MQ é acreditar em uma *interpretação* da MQ. Então como podemos resolver esse assunto, considerando que a MQ não determina sua própria interpretação? Tome esta re-esquematização do raciocínio mencionado acima sobre o argumento dos milagres, empregando as caracterizações apresentadas na seção anterior. Primeiramente:

1. A $MQ_C$ é bem-sucedida
2. A verdade da $MQ_C$ é a melhor explicação para esse sucesso.
∴ A $MQ_C$ é verdadeira.

Contudo, também poderíamos apresentar o argumento assim:

1. A $MQ_R$ é bem-sucedida
2. A verdade da $MQ_R$ é a melhor explicação para esse sucesso.
∴ A $MQ_R$ é verdadeira.

O mesmo poderia ser dito da $MQ_V$, mas vamos reduzir nossos exemplos a essas duas situações. Como a $MQ_C$ e a $MQ_R$ são empiricamente bem-sucedidas, seu sucesso parece não ter papel algum na escolha de interpretações, fazendo com que o argumento dos milagres perca sua força. E agora o amargor começa a aparecer. Apresentado dessa maneira, o argumento dos milagres deve, de alguma forma, *responder* ao argumento da subdeterminação. De qualquer modo, o argumento dos milagres parece ser de pouca ajuda para o realismo científico. Ruetsche eloquentemente coloca a situação da seguinte forma:

> Mesmo que — talvez particularmente porque — várias [interpretações] rivais estão disponíveis, existem várias possibilidades para minar o argumento dos milagres. [...] os critérios pelos quais a realista espera selecionar a interpretação vencedora falham em escolher uma dessas alternativas. (Ruetsche, 2018, p. 298)[4]

3No original: "What a realist believes when she believes a theory *T* is an *interpretation of T*, an account of what the worlds possible according to *T* are like. [...] An interpretation of QM tells the realist about QM what she believes when she believes QM".

Até o presente, então, não temos como saber qual é a dinâmica correta da MQ. Algumas concorrentes: equação determinística de continuidade ($MQ_V$), colapso estocástico ($MQ_C$), evolução determinística da equação de Schrödinger ($MQ_R$). Mas e quanto ao componente ontológico sobre o qual vínhamos falando? A subdeterminação da teoria pelos dados está diretamente relacionada, entre outras coisas, com um problema ontológico: aquilo que existe, segundo a MQ, depende da interpretação que se adota para a teoria. Havendo diferentes interpretações, todas dando conta dos mesmos dados, mas postulando ontologias distintas, não temos como escolher, com base apenas nas evidências, qual a ontologia do mundo, segundo a MQ.

Para não terminarmos essa discussão com um amargor tão forte, um copo d'água: convém observar que, apesar das dificuldades envolvidas, há uma possível saída para realistas nesse campo. Trata-se de observar que, pelo menos no que diz respeito à MQ, a discussão acerca da interpretação mais apropriada está diretamente relacionada com a formulação da dinâmica da teoria. Ao avançarmos na investigação, pode ocorrer que apenas uma dentre as diferentes opções se mostre adequada. Ou seja, esse não parece ser um problema exclusivo para realistas dos departamentos de filosofia, mas também para as pessoas dos departamentos de física que tratam de MQ. O problema, assim, deixa de ser um problema filosófico apenas, e passa a depender do próprio avanço da ciência. Conforme Esfeld salienta, há uma relação bastante próxima entre a formulação de uma teoria e a adoção de uma ontologia correspondente:

> […] uma teoria física deve (i) detalhar uma ontologia sobre o que há na natureza de acordo com a teoria (ii) fornecer uma dinâmica para os elementos da ontologia e (iii) deduzir estatísticas de resultados de medida a partir da ontologia e da dinâmica ao tratar de interações de medição dentro da ontologia e da dinâmica; para o fazer, ontologia e dinâmica devem estar ligados com uma medida de probabilidade apropriada. Assim, a questão é: qual é a lei que descreve os processos individuais que ocorrem na natureza (dinâmica) e quais são as entidades que formam esses processos individuais (ontologia)? (Esfeld, 2019, p. 222)[5]

Desde que a ontologia esteja associada a uma interpretação da MQ, as duas coisas andam juntas: ontologia e ciência. Em algum sentido do termo, essa é uma concepção "naturalizada" da ontologia, de modo que nossa concepção acerca daquilo que existe deva ser "lida" de cada interpretação. Mais um retrogosto: não só não podemos decidir entre *tipos* de solução ao problema da medição, mas também não podemos decidir entre *várias* interpretações dentro de cada tipo. Um mapa bastante completo foi oferecido por Pessoa Junior (2006), mas continuaremos na nossa simplificação com somente duas "famílias" de interpretações. Pode-se associar à $MQ_C$ a ideia de que a consciência humana causa o colapso (cf. de Barros e Oas, 2017; de Barros e Montemayor, 2019), e também à ideia de que a redução espontânea ocorra com *partículas* ou com *flashes* (cf. Allori, 2013). Da mesma forma, a $MQ_R$ pode ser associada com a existência de multiversos (cf. Wilson, 2020), mas também com a existência de um único universo (cf. Barrett, 2011; Conroy, 2012). Portanto, é difícil determinar a ontologia *mesmo que suponhamos ser capazes de determinar a teoria*.

4No original: "Even if — maybe particularly because — a variety of contenders are available, there remain several Miracles argument-undermining possibilities. [...] the criteria by which the realist hopes to select the winning interpretation fail to single any such out."

5No original: "[...] a physical theory has to (i) spell out an ontology of what there is in nature according to the theory, (ii) provide a dynamics for the elements of the ontology and (iii) deduce measurement outcome statistics from the ontology and dynamics by treating measurement interactions within the ontology and dynamics; in order to do so, the ontology and dynamics have to be linked with an appropriate probability measure. Thus, the question is: What is the law that describes the individual processes that occur in nature (dynamics) and what are the entities that make up these individual processes (ontology)?"

Todavia, apesar das dificuldades, realistas, de modo geral, não parecem se intimidar. A crença no sucesso da ciência parece ser um bom indicativo de que há algo razoável nessa abordagem. Para quem mantém sua posição, um novo aspecto do desafio surge. Veremos na próxima seção.

## 4. Uma camada metafísica

### 4.1. Ontologia e metafísica

A tese básica por trás do realismo, apesar das dificuldades que aparecem em seu caminho, parece, de fato atrativa: a ciência nos diz como é o mundo. Afinal, se a ciência não fizer o papel de nos dar uma descrição confiável (ao menos próxima da verdade), quem ou o que fará? Mas sua atratividade é também seu maior problema: *como* é o mundo segundo a descrição científica? Isto é, com qual imagem de mundo estamos nos comprometendo ao sustentarmos que a ciência diz como é o mundo? Aparentemente, não bastaria, por exemplo, repetirmos alguns aspectos centrais das teorias científicas para responder isso. Parafraseando French (2018, p. 394), suponha que alguém se declare ser realista sobre a MQ, nos apresente uma equação fundamental da MQ padrão, como a *equação de Schrödinger*, e diga: "...e é *assim* que a MQ nos diz como é o mundo". Algumas pessoas não ficariam satisfeitas com isso, com um sentimento (justificável) de que "dizer como o mundo é" requer algo *além*.

Conforme já vimos, Esfeld destaca como uma das tarefas de uma teoria física a estipulação de uma ontologia, de uma mobília para o mundo. A MQ padrão nos diz que existem elétrons, por exemplo, e interpretações mais específicas postulam a existência de uma pluralidade de mundos, de consciência, dentre outros, dependendo da interpretação. Bem, isso é um passo na direção de se fornecer uma ontologia. Mas isso basta? Há realistas que vão além, e acreditam que uma teoria precisa de mais do que povoar o mundo com entidades, se ela vai nos dar uma imagem apropriada da realidade. Trata-se de complementar a descrição científica da realidade com uma camada metafísica. Para que a discussão fique clara, vale notar que estivemos tratando do problema relacionado ao realismo científico como um problema envolvendo a 'ontologia' das nossas teorias científicas precisamente no sentido da pergunta tradicional quineana acerca daquilo que existe, da mobília do mundo (ver também os Capítulos 3 e 4 deste volume). Metafísica, em nossa acepção, possui um sentido mais geral. Assim, quando realistas pedem por uma descrição metafísica da realidade associada à ontologia, o pedido é no sentido de complementar de algum modo a descrição que a ciência nos dá. Para que isso fique mais claro, vale a pena reforçar essa distinção entre ontologia e metafísica. Conforme Hofweber esclarece:

> Na metafísica, queremos descobrir como a realidade é de um modo geral.

> Uma parte disso será descobrir as coisas ou o estofo que são partes da realidade. Outra parte da metafísica será descobrir o que essas coisas, ou esse estofo, são, de modo geral. A ontologia, nessa abordagem bem padrão da metafísica, é a primeira parte do projeto, i.e. é a parte da metafísica que busca descobrir quais coisas formam a realidade. Outras partes da metafísica elaboram sobre a ontologia e vão além dela, mas a ontologia é central para elas, […] A ontologia é geralmente conduzida perguntando-se questões sobre o que há ou o que existe. (Hofweber, 2016, p. 13)[6]

Nesse sentido, conforme estivemos discutindo, esse "algo além" que se exige da posição realista é uma metafísica. Então, pra ficar sobre a parte sem (tantas) controvérsias, alguém poderia dizer que, segundo a MQ, há objetos quânticos, como elétrons. Muito bem. Mantendo nosso exemplo, o que podemos dizer sobre o elétron, *sem repetirmos o que a física já nos diz*? Afinal, se repetirmos as partes relevantes da física, continuamos no mesmo lugar — sem o "algo além", sem uma imagem mais completa. Aqui cabe uma distinção, motivada por Magnus (2012), entre um realismo "profundo" e um realismo chamado por French (2018) de "raso". Um realismo que repete a descrição física seria o realismo raso, que se recusa a investigar questões metafísicas. Realistas do tipo *raso* seriam realistas que aceitariam as entidades postuladas pela teoria, mas se recusariam a ir além do que a ciência já diz. Realistas desse tipo despendem seus esforços para responder os argumentos antirrealistas apresentados na seção 2. Todavia, para outros realistas, isso não é profundo o suficiente. Para Chakravartty (2007, p. 26), "Não se pode apreciar completamente o que significa ser realista até que se tenha uma imagem clara acerca daquilo que estamos sendo convidados a ser realistas".[7] Fazer isso requer que se complemente a descrição científica com uma teoria metafísica, que é o que caracteriza o realismo "profundo". Quem se recusa a perceber isso não pode se declarar realista no sentido estrito do termo.

French (2014, p. 50), por exemplo, alega que o realismo profundo é o único realismo científico possível: "[a]s pessoas que rejeitam essa necessidade são ou empiristas no armário ou realistas 'ersatz'".[8] No entanto, vale notar que a camada metafísica é desejável também para alguns tipos de antirrealismo, como o empirismo construtivo. Para van Fraassen (1991, p. 242), a "questão de interpretação por excelência" é dizer como é o mundo, supondo que a teoria em questão seja verdadeira: "para entender uma teoria científica, precisamos ver como o mundo poderia ser, da maneira que a teoria diz que é". Então antirrealistas interessariam-se, a princípio, pelo desafio de Chakravartty, mesmo sem se comprometerem com a correspondência entre *o perfil metafísico de uma entidade em uma teoria* e *o mundo* — muito embora existam divergências mesmo dentro do empirismo construtivo; Bueno (2019, p. 8), por exemplo, recusa-se a "reificar os postulados na ontologia de alguém"[9] com um perfil metafísico. Seja como for, aqui podemos nos perguntar: como exatamente podemos obter essa imagem clara? Novamente, French ao resgate:

> Uma resposta simples seria: através da física, que nos dá uma imagem do mundo como incluindo partículas, por exemplo. Mas essa imagem é clara o suficiente? Considere a questão adicional, aparentemente óbvia, acerca de se

6No original: "In metaphysics we want to find out what reality is like in a general way. One part of this will be to find out what the things or the stuff are that are part of reality. Another part of metaphysics will be to find out what these things, or this stuff, are like in general ways. Ontology, on this quite standard approach to metaphysics, is the first part of this project, i.e. it is the part of metaphysics that tries to find out what things make up reality. [...] Ontology is generally carried out by asking questions about what there is or what exists."
7No original: "One cannot fully appreciate what it might mean to be a realist until one has a clear picture of what one is being invited to be a realist about."
8No original: "Those who reject any such need are either closet empiricists or 'ersatz' realists".
9No original: "reify what is posited in one's ontology".

> essas partículas são objetos individuais, assim como são as cadeiras, mesas, ou pessoas. Ao responder essas questões, precisamos fornecer, eu sustento, ou pelo menos aludir a, uma metafísica da individualidade apropriada, e isso exemplifica a afirmação geral de que para obter a imagem clara de Chakravartty e assim alcançar uma compreensão realista apropriada nós precisamos fornecer uma metafísica apropriada. (French, 2014, p. 48)[10]

Assim, a ideia é que mesmo que pudéssemos resolver as questões científicas e ontológicas, talvez ainda teríamos um último problema para resolver antes de fazermos as pazes com o realismo científico: para que tenhamos um realismo científico *legítimo*, precisamos atribuir uma camada metafísica aos resultados da ciência. *Isso* nos daria a tão almejada imagem clara. Com isso, a questão que motiva esse tipo de investigação é a seguinte: uma vez que admitimos que uma camada metafísica é necessária para o realismo científico, qual seria a *espessura* dessa camada? Isto é, o quanto de metafísica devemos admitir no nosso realismo?

Como estamos discutindo, uma das formas de se responder a esta questão é buscando extrair essa camada da própria ciência. Ela seria tão espessa quanto a ciência relevante para o caso nos diz que ela é. O problema com essa estratégia, comumente tratada como 'naturalismo em metafísica', ou 'metafísica científica', é que a ciência parece não nos dizer absolutamente nada sobre sua metafísica. O metafísico científico fica sem meios para buscar extrair da ciência qualquer informação metafísica a partir da ciência. De fato, uma teoria como a MQ, por exemplo, trata de elétrons e do comportamento de entidades microscópicas. Não trata da caracterização metafísica dessas entidades. Nesse sentido, proponentes das metafísicas científicas correm o risco de serem realistas apenas no sentido raso.

Uma forma de se evitar essas dificuldades sem recair no realismo raso seria apelar para as teorias metafísicas já existentes. Por exemplo, se por algum motivo aceitarmos que a formulação correta da MQ é aquela que envolve a admissão de uma consciência fora da descrição física que provoca o colapso da função de onda, a questão extra por uma imagem clara exigiria, entre outras, uma caracterização metafísica dessa consciência. Para tanto, podemos tentar "vestir" metafisicamente essa consciência com alguma forma de dualismo já presente na tradição filosófica ou, então, nos arriscarmos com uma roupagem metafisicamente mais econômica, tentando manter um papel para a consciência, mas a reduzindo a algum outro tipo de entidade (a discussão sobre esse caso pode ser vista em Arroyo e Arenhart, 2019). Essa tentativa de utilizar conceitos metafísicos já existentes para complementar a descrição científica é o que se chama atualmente de *abordagem Viking*, ou a *abordagem da metafísica como uma caixa de ferramentas para a ciência* (cf. French, 2014). Segundo essa abordagem, a metafísica, como vimos, aparece como um ingrediente adicional à descrição científica, e colabora para uma imagem mais clara acerca daquilo que estamos sendo realistas.

## 4.2. Um exemplo, e a subdeterminação contra-ataca

Um exemplo bastante comum de como se acrescenta essa camada metafísica pode ser apresentado quando consideramos a discussão acerca do estatuto das entidades da MQ (veja uma discussão detalhada em French e Krause, 2006). São indivíduos, como os objetos encontrados em nosso cotidiano (como gatos, pedras, cadeiras)? Ou falham nesse quesito? Supondo que as partículas com as quais a teoria quântica trata possam ser consideradas como objetos, algo similares aos objetos do quotidiano (uma suposição perigosa, mas que não podemos discutir com todos os detalhes aqui), a pergunta natural que surge para realistas, que vão acreditar na existência dessas entidades é: o que são, exatamente, essas entidades? O que as torna exatamente aquilo que elas são? Há algo fazendo esse papel? (para os conceitos fundamentais envolvendo a disputa metafísica sobre identidade e individualidade, ver French e Krause, 2006, cap. 1; Arenhart e Krause, 2012).

10No original: "A simple answer would be, through physics which gives us a certain picture of the world as including particles, for example. But is this clear enough? Consider the further, but apparently obvious, question, are these particles individual objects, like chairs, tables, or people are? In answering this question we need to supply, I maintain, or at least allude to, an appropriate metaphysics of individuality and this exemplifies the general claim that in order to obtain Chakravartty's clear picture and hence obtain an appropriate realist understanding we need to provide an appropriate metaphysics."

Essa é uma discussão bastante importante na MQ, principalmente pelo fato de que alguns dos fundadores da teoria parecem ter expressado a opinião de que partículas quânticas apresentam comportamento distinto das partículas clássicas precisamente nesse aspecto envolvendo a sua individualidade. Schrödinger abre um de seus ensaios sobre o assunto do seguinte modo:

> Este ensaio trata da partícula elementar, mais particularmente, de uma característica que esse conceito adquiriu — ou antes perdeu — na mecânica quântica. Quero dizer isso: que uma partícula elementar não é um indivíduo; ela não pode ser identificada, ela não possui “igualdade”. […] Em linguagem técnica isso é explicado ao se dizer que as partículas “obedecem” a uma nova estatística, seja a estatística de Bose-Einstein, seja a estatística de Fermi-Dirac. A implicação, longe de ser óbvia, é que o epíteto insuspeito “isso” não é apropriadamente aplicável, digamos, a um elétron, exceto com cuidado, em um sentido restrito, e algumas vezes simplesmente não se aplica. (Schrödinger, 1998, p. 197)[11]

Em termos não muito técnicos, podemos ilustrar o que nos dizem as estatísticas com os seguintes exemplos (veja também Arenhart e Krause, 2012). Suponha que desejamos distribuir duas partículas idênticas em todas as suas propriedades (indiscerníveis), rotuladas 1 e 2, em dois estados possíveis, A e B. Escreveremos A(1) para dizer que a partícula 1 se encontra no estado A, e de modo similar para outras configurações. Se 1 e 2 rotulam partículas clássicas, e assumindo-se que todas as distribuições entre estados são equiprováveis (i.e. que não há um estado privilegiado), temos quatro possibilidades: i) A(1)A(2) (tanto 1 quanto 2 estão em A); ii) A(1)B(2); iii) A(2)B(1) e iv) B(1)B(2). Note que ii) e iii) diferem apenas por uma permutação dos rótulos das partículas. Ainda assim, contam como casos diferentes: faz diferença se é 1 que está em A e 2 em B (caso ii), ou se é 2 que está em A e 1 em B (caso iii).

Para partículas *quânticas* indiscerníveis, as possibilidades de distribuição são diferentes (e é sobre isso que Schrödinger estava falando). Para partículas chamadas bósons (que obedecem a estatística Bose-Einstein), temos apenas três casos: i) A(1)A(2), ii) A(*)B(*) e iii) B(1)B(2). Ou seja, além de poderem estar no mesmo estado (A ou B), temos uma situação ii) em que temos uma partícula em A e outra em B. O asterisco está presente precisamente porque não distinguimos, como no caso clássico, situações em que é 1 que está em A e 2 em B, de situações em que 2 está em A e 1 em B. *Permutar os rótulos das partículas não origina uma situação diferente*. Algo similar ocorre com *férmions*: férmions não podem estar no mesmo estado, de modo que as situações i) e iii) dos bósons são impossíveis para eles. Nos resta apenas o caso i) A(*)B(*), ou seja, o caso em que uma partícula está em A e outra em B (essa é a única distribuição possível, e novamente, não há diferença em se permutar os rótulos das partículas).

11No original: “This essay deals with the elementary particle, more particularly with a certain feature that this concept has acquired — or rather lost — in quantum mechanics. I mean this : That the elementary particle is not an individual; it cannot be identified, it lacks “sameness.” [...] In technical language it is covered by saying that the particles “obey” new-fangled statistics, either Einstein-Bose or Fermi-Dirac statistics. The implication, far from obvious, is that the unsuspected epithet “this” is not quite properly applicable to, say, an electron, except with caution, in a restricted sense, and sometimes not at all.”

Como explicar que a situação clássica faça uma distinção entre duas situações (chamadas de ii) e iii) no caso clássico) que a descrição quântica não faz? Schrödinger, como vimos, faz um apelo a uma noção metafísica para tanto: ao contrário das partículas clássicas, que são indivíduos, as partículas quânticas perderam essa característica, de modo que o resultado é serem insensíveis a permutações. Em outras palavras: apesar de terem as mesmas propriedades, partículas clássicas possuem individualidade (há algo que faz com que a partícula 1 seja ela mesma, e não idêntica à partícula 2), e que explicaria a diferença entre os casos ii) e iii) na contagem clássica. Ora, esse algo a mais, a individualidade, está ausente na MQ e, desse modo, os casos de permutações não são contados como distintos. Na MQ não há nada mais para fazer uma diferença, de modo que as permutações não originam situações diferentes.

Isso daria um primeiro passo na busca por uma roupagem metafísica para as entidades quânticas. Ao aceitarmos que existem partículas quânticas, estamos, segundo essa sugestão, aceitando que existem também entidades sem individualidade, e essa seria sua *roupagem metafísica*. Mas a questão que deve estar lhe perturbando agora é: *o que, exatamente são não-indivíduos*? Uma maneira de se entender essa questão consiste em novamente seguir Schrödinger, que em outro lugar sugeriu a seguinte imagem para os não-indivíduos:

> Eu volto a enfatizar isso e rogo que você acredite: não é uma questão de sermos capazes de asserir a identidade em alguns casos, e não sermos capazes de asserir a identidade em outros. Está além da dúvida que a questão da "igualdade", da identidade, real e verdadeiramente não possui significado. (Schrödinger, 1996, pp. 121–122)[12]

A situação descrita por ele aqui é a seguinte: ao se detectar uma partícula em um instante $t_1$, e logo em seguida, ao se detectar um traço de partícula em outro instante subsequente $t_2$, mesmo que pareça óbvio que se trata da mesma partícula traçando uma trajetória, isso não pode ser afirmado segundo a MQ. Isso porque a noção de 'mesma partícula' perde o sentido em MQ. Devemos resistir a essa atribuição natural de identidade que faríamos no caso clássico. Não há sentido em se afirmar identidade e diferença de partículas na maioria das situações em MQ. Em outras palavras, a noção de não-indivíduo foi formulada de modo a significar que a noção de identidade não faz sentido para essas entidades; não faz sentido dizer que são iguais ou diferentes (ver French e Krause, 2006, cap. 4 e cap. 7 para a articulação padrão dessa concepção, e Arenhart e Krause, 2012 para mais discussão). As partículas clássicas, por outro lado, seriam indivíduos por possuírem identidade.

Isso parece avançar na direção certa, e realistas acerca da MQ começam a preencher as lacunas de compreensão das entidades acerca das quais são realistas. Todavia, isso nos dá uma imagem mais clara das entidades quânticas? E como justificar nossa crença em não-indivíduos? Basta um apelo ao funcionamento das estatísticas? A verdade é que o sucesso empírico da MQ não parece se estender ao domínio da metafísica aqui, apesar da autoridade de Schrödinger parecer nos indicar o contrário. De fato, a correspondente metafísica da não-individualidade não recebe confirmação a partir do sucesso empírico da MQ. Ela parece ser uma dose extra de teorização acrescentada à teoria, mas não é um dos fatores explicativos que fundamentam o funcionamento da teoria. Em particular, não é exigida para se explicar o funcionamento das estatísticas quânticas (pelo contrário, ao eliminar a identidade, ela parece revisar o vocabulário com o qual formulamos a teoria, mas isso já é outro assunto).

12No original: "And I beg to emphasize this and I beg you to believe it: It is not a question of our being able to ascertain the identity in some instances and not being able to do so in others. It is beyond doubt that the question of 'sameness', of identity, really and truly has no meaning."

Isso fica mais claro se trouxermos um pouco mais de detalhes para a nossa discussão. Se entendermos que o problema metafísico da individualidade consiste em se explicar o que é uma entidade, em oposição a todas as outras (por exemplo, o que faz de Sócrates o indivíduo que ele é, e não qualquer outro), encontramos na literatura diferentes teorias sobre o que confere individualidade a um indivíduo (cf. French e Krause, 2006, cap. 1). Duas abordagens bastante tradicionais são a chamada *teoria dos feixes de propriedades*, segundo a qual um indivíduo é completamente definido ou caracterizado pela coleção de propriedades que ele possui (uma teoria que, aparentemente, impede que dois indivíduos possuam as mesmas propriedades), e as *teorias da individualidade transcendental*, que identifica o princípio de individualidade ou com um *substrato* (um ingrediente extra que compõe o indivíduo, para além de suas propriedades), ou com uma propriedade não-qualitativa, exclusiva ao indivíduo (conhecida como *hacceidade* ou *essência individual*). Tipicamente, alega-se que a indiscernibilidade das entidades quânticas torna a teoria dos feixes inadequada para se conferir individualidade aos quanta. Ainda, as hacceidades costumam ser entendidas como a propriedade de ser idêntico a si mesmo. Assim, a hacceidade de Sócrates, o que lhe confere individualidade segundo essa abordagem, é a propriedade de *ser idêntico a Sócrates,* que certamente seria instanciada apenas por Sócrates. Ora, se a teoria de feixes está fora da jogada devido à indiscernibilidade, e Schrödinger sugeriu que a identidade não faz sentido para as entidades quânticas, a não-individualidade quântica é caracterizada *metafisicamente* como a ausência de hacceidades para as partículas quânticas (essa é, muito de perto, a posição sugerida por French e Krause, 2006, cap. 4; ver também Arenhart, 2017). Indivíduos possuem hacceidade, e não-indivíduos não possuem. Isso resolve o nosso problema, não?

Na verdade, não. Nada nos obriga a aceitar que a identidade *não faz sentido* para as entidades quânticas. Em outras palavras, a teoria quântica não nos impõe a teoria de que entidades quânticas são não-indivíduos precisamente no sentido de não possuírem hacceidade. Mais do que isso: nada, na MQ, proíbe que essas entidades sejam individuadas por um substrato! De fato, a teoria é compatível com a existência de substratos individuando as partículas (lembre: segundo a teoria do substrato, duas partículas podem ser indiscerníveis segundo suas propriedades, mas ainda assim contarem como indivíduos diferentes por conta de seus substratos). Algo similar pode ser dito até mesmo para hacceidades. Assim, há diferentes interpretações metafísicas da MQ nas quais as entidades em questão *são indivíduos,* apesar do que sugere Schrödinger (e French e Krause, 2006, cap. 4, reconhecem isso, certamente). O que temos é simplesmente uma nova situação, em que a MQ, além de estar subdeterminada pelos dados, subdetermina sua metafísica da individualidade e não-individualidade (cf. French e Krause, 2006, cap. 4, e Arenhart, 2017 para ainda mais subdeterminação). Isso torna mais difícil a vida de realistas que desejam acrescentar uma camada de metafísica à descrição científica, dado que são vítimas da subdeterminação duas vezes! Lembrando da distinção proposta por Muller (2011), além da "primeira tese" da subdeterminação, a subdeterminação metafísica seria a "segunda tese" da subdeterminação.

Note: esse argumento ameaça apenas quem deseja extrair algum tipo de apoio da ciência para a sua metafísica, e que acredita que necessita dessa camada extra para ter uma descrição apropriada da realidade que mereça ser chamada de realismo (ou seja, quem aceita o desafio de Chakravartty). Para realistas que recusam essa tarefa adicional, todavia, como já vimos, permanece a acusação de serem empiristas que não se assumiram enquanto tal, já que não se aprofundam o suficiente. Aqui devemos ressaltar novamente que 'realismo', no contexto deste texto, significa 'realismo científico', e 'empirismo', significa 'empirismo construtivo e suas variantes'. De modo geral, o empirismo construtivo é um antirrealismo

científico, apesar de ser um realismo acerca de observáveis. As escolhas não são fáceis. Além disso, esse é um simples exemplo concernente à individualidade. Outros casos podem se multiplicar, tratando da natureza do espaço, do tempo, das propriedades, da composição, entre outros. Realistas que desejam uma imagem clara acerca disso, certamente possuem um amplo leque de escolhas, e é aqui que está a sua infelicidade: subdeterminação.

#### 4.2.1.Realismo estrutural: um último recurso

French (2014) argumenta que um tipo específico de realismo pode triunfar sobre os argumentos antirrealistas, caso o realismo seja aceito apenas com relação às estruturas das teorias científicas. Essa posição é o *realismo estrutural* que, supostamente, pode responder produtivamente aos desafios impostos pela subdeterminação. A posição advogada por French é o realismo estrutural ontológico, e afirma que somente as *estruturas existem*. É uma tese que se encontra em oposição a realismos estruturais menos radicais, como o realismo estrutural *epistêmico* (que diz que podemos *conhecer* as estruturas, somente, cf. Worrall, 1989). Um argumento tipicamente empregado em defesa do realismo estrutural ontológico foi esquematizado por Branding e Skiles (2012, pp. 100–101), mas modificamos seu argumento para atender aos propósitos das subdeterminações advindas das interpretações da MQ — que foi o exemplo com o qual estivemos trabalhando até aqui.

- *Premissa 1:* Realistas orientados por objetos carregam comprometimento ontológico com objetos, e esses objetos podem variar: (i) existe um compromisso com uma consciência causal no $MQ_C$ e com mundos divididos no caso do $MQ_R$ e (ii) não há fato estabelecido sobre o assunto sobre qual desses tipos de objetos existe.
- *Premissa 2:* Se a premissa 1 for o caso, a adoção do realismo orientado por objetos implica em um compromisso com a expectativa de que as melhores teorias descrevem com precisão quais objetos existem.
- *Premissa 3:* As melhores teorias, no entanto, falham em oferecer um relato de quais objetos realmente existem: a ontologia, conforme apresentada pelas melhores teorias, é ontologicamente subdeterminada.
- *Conclusão 1:* Portanto, o realismo orientado por objetos é (provavelmente) falso.

De acordo com o realismo estrutural, o argumento acima pode ser adaptado da seguinte maneira.

- *Premissa 4:* Se o realismo estrutural é verdadeiro, então nossas melhores teorias não são infectadas pela subdeterminação ontológica.
- *Conclusão 2:* Assim, o realismo estrutural é preferível a um realismo orientado por objetos.

Outra vantagem é que, alega-se, o realismo estrutural também supera, como um bônus, o problema da subdeterminação metafísica. Ao evitar compromisso com objetos, a questão acerca de sua identidade e individualidade simplesmente não aparece, e evitamos as dificuldades discutidas a pouco. Essa foi a sugestão de Ladyman ao propor que o realismo estrutural é uma forma de realismo que evita deixar questões importantes indeterminadas:

> Devemos reconhecer a falha de nossas melhores teorias em determinar até mesmo a característica ontológica mais fundamental das entidades que elas supostamente descrevem. Uma forma de realismo que recomenda crença na existência de entidades que possuem um estatuto metafísico tão ambíguo é

> um realismo *ersatz*. O que é exigido é uma mudança para uma base ontológica inteiramente diferente, uma para a qual questões de individualidade simplesmente não surgem. (Ladyman, 1998, pp. 419–420)[13]

Ladyman está falando especificamente sobre a subdeterminação metafísica. A sugestão é a de que um realismo que sofre com subdeterminação (mesmo a metafísica) não é um realismo legítimo. O realismo estrutural seria uma opção para evitar essas dificuldades. Mas será que mudar para uma ontologia de estruturas evita as dificuldades?

O realismo estrutural ontológico enfrenta problemas tanto com o realismo raso quanto com o profundo. Da parte "rasa", não é absolutamente óbvio como o realismo estrutural responde à subdeterminação. Como Esfeld (2012) argumenta, o realismo estrutural "[...] não é uma interpretação da MQ, em adição às interpretações do tipo muitos mundos, às interpretações do tipo colapso ou às interpretações do tipo variáveis ocultas",[14] mas está ligada a cada uma dessas interpretações específicas da MQ. Assim, quando o realismo estrutural propõe que sejamos realistas sobre a MQ, qual realismo estaria implícito? Isto é, sobre a $MQ_V$, $MQ_C$, $MQ_R$, ou, ainda, algo além? Essas perguntas devem ser respondidas quanto à interpretação correta da MQ quando se adota uma posição realista sobre a estrutura de tal interpretação. Um horizonte de respostas não é sequer claro:

> Uma realista estrutural pode reagir à subdeterminação da interpretação pela teoria quântica com as esperanças de que (i) exista alguma estrutura que todas essas interpretações compartilhem e que (ii) a identificação dessa estrutura constitua uma especificação do conteúdo do realismo estrutural sobre a MQ. O ponto (ii) atenuaria a preocupação em nível de satélite sobre o realismo estrutural de que seu conteúdo é irremediavelmente indeterminado. Infelizmente, mesmo limitando a atenção às interpretações rivais, não é absolutamente claro que elas tenham qualquer estrutura em comum, *que seja do interesse do realismo*. (Ruetsche, 2018, p. 300, ênfase original)[15]

Portanto, qualquer descrição realista estrutural sobre a MQ parece conter o ônus da prova para resolver a subdeterminação interpretativa. No entanto, como décadas de debate sem consenso sobre os fundamentos da MQ atestam, isso não é de modo algum uma tarefa fácil. Como disse Callender (2020, p. 75) em um tom quase apocalíptico, "Em suma, temos uma séria subdeterminação científica. O pesadelo dos realistas científicos é real".[16]

Para tornar tudo ainda mais difícil, não é óbvio que ao adotarmos o realismo estrutural conseguimos escapar da subdeterminação metafísica. O fato de evitarmos compromissos com objetos e aceitarmos compromissos com estruturas simplesmente muda o tipo de entidade que vai precisar de uma roupagem metafísica, mas a necessidade de tal roupagem permanece para que não se tenha um empirismo de armário. O desafio de Chakravartty, proposto por French, afirma que um realismo sobre algo (digamos, estruturas) é um realismo que carece de conteúdo, *a menos que* uma caracterização metafísica sobre esse algo seja oferecida. Como reconhecido por Arenhart e Bueno (2015), até então não há perfis metafísicos para entendermos, metafisicamente, o que pode ser uma estrutura. Portanto, segundo os próprios critérios de French (2014, p. 50), o "realismo estrutural" parece não ser um realismo sobre estruturas, já que falha em responder ao desafio de Chakravartty.

13No original: "We need to recognize the failure of our best theories to determine even the most fundamental ontological characteristic of the purported entities they feature. It is an ersatz form of realism that recommends belief in the existence of entities that have such ambiguous metaphysical status. What is required is a shift to a different ontological basis altogether, one for which questions of individuality simply do not arise."

14No original: "OSR [ontological structural realism] is not an interpretation of QM [quantum mechanics] in addition to many worlds-type interpretations, collapse-type interpretations, or hidden variable-type interpretations".

15No original: "A structural realist might react to the underdetermination of interpretation by quantum theory with the judo-like hopes that (i) there is some structure all these interpretations share and that (ii) identifying that structure constitutes a specification of the content of structural realism about QM. Point (ii) would assuage the satellite-level concern about structural realism that its content is hopelessly indeterminate. Alas, even confining attention to the contender interpretations, it is not at all clear they have in common any structure *of interest for realism.*"

16No original: "In sum, we have serious scientific underdetermination. The nightmare of scientific realists is real."

### 4.3.Nem tudo está perdido?

Mas nem tudo está completamente perdido na relação entre ciência e metafísica. Enquanto o metafísico científico fica sem recursos para delinear sua metafísica (caso o queira fazer), e o partidário da abordagem Viking se torna vítima da segunda tese da subdeterminação, há, pelo menos a princípio, uma terceira via (para mais argumentos acerca da subdeterminação da metafísica na abordagem Viking, ver Arroyo 2020).

Conforme você pode ter percebido, em nossa discussão sobre o caso da individualidade em MQ, dissemos acima que uma teoria de feixes é incompatível com a MQ. Vamos desenvolver brevemente esse ponto.

Para as pessoas que aceitam que a MQ trata de objetos, e que esses objetos devem ter um perfil metafísico para que sejam entendidos mais claramente, *uma das opções que não está disponível* é a teoria de feixes. Recorde: a teoria de feixes indica que um indivíduo é caracterizado pela coleção de suas propriedades. Um princípio que é associado com essa teoria é o chamado *Princípio da Identidade dos Indiscerníveis* (PII): se dois objetos são qualitativamente idênticos (possuem, igualmente, todas as propriedades, isto é, *as mesmas* propriedades), então eles são numericamente idênticos (são apenas um e o mesmo objeto). Aqui, existem algumas sutilezas que merecem um comentário. O que exatamente queremos dizer com 'ter as mesmas propriedades'? Uma versão bastante exigente do PII sugere que isso significa ter as mesmas qualidades, sem envolver relações com outros indivíduos, nem envolver relações espaciais. Assim, PII, segundo essa formulação exigente, estaria sugerindo que dois objetos ocupando posições diferentes são diferentes ainda em alguma outra qualidade, não fazendo referência às posições, dado que estas não podem ser usadas para garantir a diferença numérica. Uma segunda versão do PII permite que relações possam ser empregadas. Com isso, objetos que possuem as mesmas qualidades, mas não entrem nas mesmas relações, podem ser diferenciados precisamente por estas relações. Por fim, uma terceira versão permitiria que objetos possam ser diferenciados por suas qualidades, por suas relações, ou até mesmo por suas posições no espaço. Cada uma dessas versões do PII possui seus próprios problemas, que não nos cabe discutir aqui. Todavia, não importa o quanto se esteja disposto a flexibilizar o PII, aparentemente a MQ viola todas as formas do princípio. Ou seja, os objetos quânticos podem partilhar as mesmas propriedades (em qualquer das versões para 'qualidades' que vimos), e ainda assim não serem numericamente os mesmos.

Essa violação do PII acarreta uma consequente violação da teoria de feixes. Objetos podem ser qualitativamente idênticos e, ainda assim, não serem apenas um só. Isso indicaria que a teoria de feixes, uma teoria metafísica acerca da individualidade, seria desbancada pela MQ acompanhada de uma ontologia de objetos. Note: a MQ forneceria contraexemplos para o funcionamento de uma teoria metafísica e, se tratando de uma de nossas mais bem-sucedidas teorias físicas, esse contraexemplo é muito mais convincente do que os exemplos filosóficos fornecidos até o momento (para uma discussão detalhada sobre o PII na MQ, ver French e Krause, 2006, cap. 4; há toda uma literatura mais recente questionando a falha do PII na MQ, mas a discussão nos levaria mais longe do que é desejável neste capítulo).

Com o fracasso do PII, a teoria dos feixes, aparentemente, deixa de servir como uma candidata razoável à teoria da individualidade. Qual a lição metodológica que podemos tirar dessa aparente derrota da teoria dos feixes na MQ? Ora, que apesar de ainda termos disponível uma teoria de substrato e de hacceidades, além de diversas teorias de não-indivíduos, quando se trata de estabelecer um perfil metafísico para estas entidades, ainda assim podemos contar com nossa teoria científica para *eliminar* candidatos, diminuindo a carga de subdeterminação metafísica. Essa abordagem, proposta, entre outros lugares, em Arenhart (2012, 2019), sugere que apesar de não podermos ter uma resposta positiva da ciência acerca de sua metafísica, podemos, depois de determinar uma ontologia, eliminar metafísicas incompatíveis (para uma aplicação ao caso da interpretação da mente causando colapso, ver Arroyo e Arenhart, 2019, e ver também mais discussão em Arroyo, 2020).

Por sua semelhança com a proposta de Popper para o progresso da ciência, a proposta aqui esboçada para uma interação entre ciência e metafísica se chama de método meta-Popperiano. 'Popperiano', por utilizar 'refutações', 'meta', por se tratar de uma forma de epistemologia para a metafísica (sendo assim uma forma de metametafísica). Isso seria de pouco consolo para as pessoas que esperam uma relação mais direta entre metafísica e ciência, que gostariam de ver o conteúdo metafísico ser de algum modo extraído da ciência, mas na falta de qualquer esperança de se poder realmente extrair metafísica da ciência, é uma forma de relação positiva entre metafísica e ciência. Segundo essa proposta, cabe aos metafísicos articular suas teorias de modo mais detalhado possível, para que se encaixem sobre as entidades postuladas por uma teoria, e para que se possa julgar se a metafísica pretendida realmente se enquadra na teoria científica. Com todas as esquisitices da MQ, podemos esperar que outras abordagens metafísicas possam falhar, e que o número de opções pode ser reduzido significativamente. Isso, no entanto, é trabalho para outra ocasião.

## 5.Conclusão

Conforme discutimos neste capítulo, o realismo parece ser uma posição bastante atrativa para as pessoas que consideram a ciência como nossa melhor forma de acesso ao mundo. Todavia, os desafios para se articular a posição ainda são muitos. Vamos recapitular brevemente como surgem esses desafios e como eles reaparecem na busca de uma articulação entre metafísica e ciência.

Como vimos, o realismo científico é a posição razoável a se adotar quando se acredita que nossas melhores teorias científicas estão corretas. Seu sucesso conta como um indicador de que, em certa medida, as teorias estão nos dando uma descrição correta da realidade. O argumento dos milagres busca tornar este vínculo mais explícito. Basicamente, o argumento nos auxilia a justificar nossa crença na ontologia proposta pela teoria, até mesmo (ou principalmente) quando esta ontologia envolve objetos não observáveis. Quando nos focamos na MQ, um passo a mais deve ser dado, dadas as dificuldades relacionadas com a dinâmica da teoria: é preciso interpretar a teoria, e isso envolve, em geral, postular uma ontologia extra (que pode envolver mentes, novos mundos, entre outros). A ontologia da teoria é dada por uma interpretação, e realistas acreditam no conteúdo de uma *interpretação*.

As dificuldades para se obter uma justificação para o realismo são os conhecidos argumentos da metaindução pessimista e o argumento da subdeterminação da teoria pelos dados. Basicamente, o primeiro desses argumentos nos lembra de que a MQ, assim como nossas melhores teorias do passado, podem ser substituídas num futuro breve por teorias melhores, com outras ontologias sendo propostas. Isso seria motivo suficiente para não aceitarmos a ontologia da MQ como fornecedora da resposta definitiva para a mobília do mundo. O argumento da subdeterminação sugere que sequer podemos determinar a mobília do mundo, mesmo se considerarmos as interpretações da teoria, dado que todas elas são igualmente empiricamente bem-sucedidas. Nada nos dados nos permite escolher entre uma interpretação em detrimento de outras.

A subdeterminação também volta a assombrar o realismo quando consideramos que, para ser realista, não basta apontar para uma ontologia: segundo algumas pessoas, é preciso ter uma imagem clara das entidades acerca das quais se é realista (o desafio de Chakravartty). Isso envolve (segundo as concepções em voga) dotar de um perfil metafísico as entidades postuladas pelas nossas melhores teorias, vestir metafisicamente as entidades que compõem a mobília do mundo. Vimos como isso costuma ser feito no caso da metafísica da individualidade e não-individualidade na MQ. O problema é que existem muitas formas de se fazer essa relação entre metafísica e ontologia de uma teoria, todas elas, a princípio, compatíveis com a teoria. A subdeterminação volta, mas agora acerca da metafísica. O perigo para uma forma de realismo que sofre com esse tipo de subdeterminação, conforme vimos, é o de que ao deixar tantas questões indeterminadas, não temos uma versão legítima de realismo (veja novamente Ladyman, 1998, pp. 419–420).

Uma forma de se evitar essa subdeterminação seria resistir ao apelo para que uma camada metafísica seja adicionada sobre a ontologia. Quem faz isso acaba ficando apenas com a ontologia, mas ao custo de também perder a legitimidade de seu realismo (são empiristas não-assumidos, segundo French). A sugestão do realismo estrutural ontológico é a mudança para uma nova base ontológica, ou seja, devemos esquecer os objetos (afinal, são eles que pedem uma metafísica da individualidade), e ficar apenas com uma ontologia de relações. O resultado é conhecido como realismo estrutural ontológico, a crença de que tudo o que há são estruturas ou relações. Como vimos, segundo proponentes do realismo estrutural ontológico, ao eliminar objetos, não viramos vítimas da subdeterminação da metafísica pela física, dado que não há mais objetos para que a subdeterminação se aplique. Todavia, mesmo que alguém aceite isso, ao afirmar uma ontologia de relações, a pergunta pela roupagem metafísica deve ser repetida (afinal, o desafio de Chakravartty permanece): o que são relações? Universais? Tropos? Parece que a mudança de base ontológica, apesar de tudo, não avança na discussão.

Outra forma, que discutimos também, seria entender a relação entre metafísica e realismo de maneira mais negativa. Segundo a sugestão que propomos, as teorias científicas, através de suas ontologias, não determinam uma roupagem metafísica, mas, ao menos, nos indicam quais dessas roupagens são inadequadas. Haveria uma forma de, progressivamente, eliminarmos metafísicas incompatíveis. Diante das peculiaridades da MQ, é bem provável que essa eliminação seja bastante efetiva, mas essa tarefa ainda precisa ser feita.

**Referências**